# Numerical simulation of $BOD_5$ dynamics in Igapó I lake, Londrina, Paraná, Brazil: Experimental measurement and mathematical modeling

**Neyva Maria Lopes Romeiro**[a,1], **Fernanda Bezerra Mangili**[b,2], **Ricardo Nagamine Costanzi**[c,3], **Eliandro Rodrigues Cirilo**[a,4], **Paulo Laerte Natti**[a,5]

[a]Universidade Estadual de Londrina (DMAT/UEL), CEP 86057-970 Londrina, Paraná, Brazil.
[b]Universidade Estadual de Londrina (ENGES/UEL), CEP 86057-970 Londrina, Paraná, Brazil .
[c]Universidade Tecnológica Federal do Paraná (UTFPR/LD), CEP 86036-370 Londrina, Paraná, Brazil.

[1]e-mail: nromeiro@uel.br
[2]e-mail: fermangili@gmail.com
[3]e-mail: ricardocostanzi@gmail.com
[4]e-mail: ercirilo@uel.br
[5]corresponding author e-mail: plnatti@uel.br
Tel: 55-43-33714226   Fax: 55-43-33714216

**Abstract.** The concentration of biochemical oxygen demand, $BOD_5$, was studied in order to evaluate the water quality of the Igapó I Lake, in Londrina, Paraná State, Brazil. The simulation was conducted by means of the discretization in curvilinear coordinates of the geometry of Igapó I Lake, together with finite difference and finite element methods. The evaluation of the proposed numerical model for water quality was performed by comparing the experimental values of $BOD_5$ with the numerical results. The evaluation of the model showed quantitative results compatible with the actual behavior of Igapó I Lake in relation to the simulated parameter. The qualitative analysis of the numerical simulations provided a better understanding of the dynamics of the $BOD_5$ concentration at Igapó I Lake, showing that such concentrations in the central regions of the lake have values above those allowed by Brazilian law. The results can help to guide choices by public officials, as: (i) improve the identification mechanisms of pollutant emitters on Lake Igapó I, (ii) contribute to the optimal treatment of the recovery of the polluted environment and (iii) provide a better quality of life for the regulars of the lake as well as for the residents living on the lakeside.

# Simulação numérica da dinâmica de DBO5 no lago Igapó I, Londrina, Paraná, Brasil: Medição experimental e modelagem matemática

**Resumo.** A concentração da demanda bioquímica de oxigênio, BOD5, foi estudada para avaliar a qualidade da água do lago Igapó I, em Londrina, Paraná, Brasil. A simulação foi realizada por meio da discretização da geometria do lago Igapó I em coordenadas curvilíneas, juntamente com os métodos de diferenças finitas e elementos finitos. A avaliação do modelo numérico proposto para a qualidade da água foi realizada através da comparação dos valores experimentais de DBO5 com os resultados numéricos. A avaliação do modelo mostrou resultados quantitativos do parâmetro simulado compatíveis com o comportamento real do lago Igapó I. A análise qualitativa das simulações numéricas permitiu uma melhor compreensão da dinâmica da concentração de DBO5 no lago Igapó I, mostrando que tais concentrações nas regiões centrais do lago apresentam valores acima daqueles permitidos pela legislação brasileira. Nossos resultados podem ajudar a orientar escolhas pelos órgãos públicos, como: (i) melhorar os mecanismos de identificação de emissores de poluentes no Lago Igapó I, (ii) contribuir para o tratamento ótimo da recuperação do ambiente poluído e (iii) proporcionar uma melhor qualidade de vida para os frequentadores do lago, bem como para os moradores que vivem na beira do lago.

**Palavras-chave:** Simulação numérica, coordenada curvilínea, método de diferenças finitas, método de elementos finitos, dinâmica da demanda bioquímica de oxigênio.

**Introduction**

Rivers, lakes, lagoons water quality models can be valuable tools to evaluate and manage the hydric systems water quality by generating essential information to promote changes in consumption and the development of preservationist urban policies. This study aims to present a mathematical model whose results can be adopted by public officials for the recovery of water quality in Igapó I lake in Londrina, Paraná, Brazil. It is about preservation of an environmental asset - the water - essential for a healthy quality of life and to maintain the local ecological balance, both fundamental rights so declared by the Constitution of the Federative Republic of Brazil from 1988.

In this context, the Igapó lake, which is subdivided into Igapó I, II, III e IV, has gone through an increasing process of degradation due to the population density around it and, consequently, to the increase in dejects dumped into its waters, causing pollution and environmental problems. Alternatives to preserve systems such as that of the Igapó Lake can be developed through an understanding of processes highly dependent on hydrodynamics and water quality parameters. Numerical models may help understand the self-depuration of systems such as that of the Igapó Lake as well as determine pollution control measures associated with environmental management, making possible an evaluation of the impact caused by pollutants discharged into the lake (ROMEIRO et al., 2011).

In order to simulate the effects of such a discharge, a two-dimensional horizontal model, composed of a hydrodynamic model and a transport model, is used. In this model the water flow in the discretized geometry (CIRILO; BORTOLI, 2006) is described by Navier-Stokes and pressure equations in curvilinear coordinate – the hydrodynamic model - (ROMEIRO et al., 2011), whereas the transport of the concentration of biochemical oxygen demand, $BOD_5$, is described by an advective-diffusive equation in curvilinear coordinate – the transport model - (PARDO et al., 2012).

On the numerical procedures, we use the Finite Differences Method (FDM) to discretize the Partial Differential Equations (PDEs). By the Succesive Over Relaxation Method (SOR), the algebraic equation system originating from pressure equation was resolved iteratively, generating the pressure field for the whole computational domain (CIRILO et al., 2008). The Navier-Stokes velocity components ($u$ and $v$) were obtained for the Igapó I Lake geometry by inserting the pressure field in the Navier-Stokes equations, and these equations were solved via Runge-Kutta Method of third order (ROMEIRO et al., 2011; PARDO et al., 2012; FERREIRA, V. G. et al., 2012). The MAC (Marker-and-Cell) Method was the technique used to solve the system of equations (AMSDEN; HARLOW, 1970). To transport model, the Semi-Discrete Stabilized Finite Element (SDSFE), a method that combines finite differences in time and finite elements in space (DONEA; HUERTA, 2003; ROMEIRO et al., 2007; MALTA; CASTRO, 2010), was used. The Navier-Stokes velocity field was used as entry data in the transport model.

The paper is organized as follows: Section 2 (Materials and methods) presents some relevant physical properties of the Igapó Lake. In this section are presented the flow of the tributaries, and the collection sites for measuring the concentration of biochemical oxygen demand, $BOD_5$. Section 3 (Computational model) describes the computing model in curvilinear coordinate (generalized coordinates) of the two-dimensional geometry discretization of the Igapó I Lake. The water quality model of our study, consisting of the hydrodynamic model and the transport model, are also presented in this section. At last, in Section 4 (Results and discussion), a qualitative analysis of the numerical results is conducted.

**Materials and methods**

**Physical and geographical characteristics of Igapó I Lake**

In December 1959, more than two decades after the creation of the county of Londrina, the Igapó Lake was created by the impoundment of Cambé stream, to solve drainage problems caused by a natural rocky dam. The lake is divided into Igapó I, II, III and IV, as shown by the satellite image from Google Earth (GOOGLE EARTH, 2016), Figure 1.

For being located near the central area of the city of Londrina, Igapó I Lake receives the discharge of untreated pollutants in its waters plus the discharge of pollutants from Lakes IV, III and II, which pollute Lake I. As observed in Figure 1, the water flows from Igapó II Lake into Igapó I Lake, which characterizes its entrance as tributary T1. On the left bank, there is low lying vegetation as well as an input channel, the Leme stream, referred to as tributary T2. The right bank is split into private properties and contains another input channel, referred to as tributary T3. The exit is a physical dam and the water flow is controlled by water pipes and ramps.

**Figure 1:** Igapó Lake I, II, III e IV and tributaries T1, T2 and T3 which flows into Igapó Lake I.

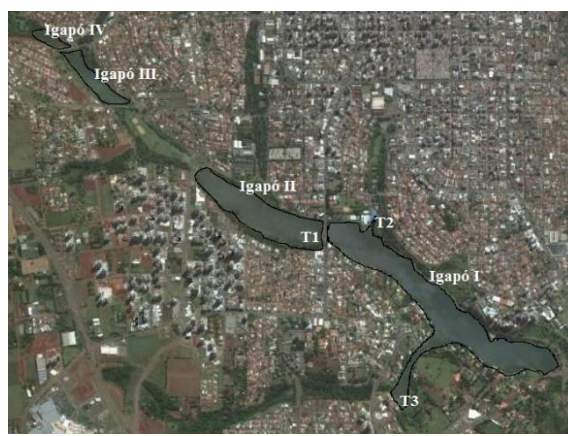

**Source:** Google Earth, 2016 (adapted).

**Tributaries flow measurement**

Distinct methodologies were adopted to determine the flow of tributaries T1, T2 and T3, since they have different characteristics. Figure 1 (GOOGLE EARTH, 2016) shows the physical domain geometry of Igapó I Lake and the locations where the flow of T1, T2 and T3 was measured.

Tributary T1, the Igapó II Lake, runs into the Igapó I Lake through a stair-like cascade. T1 flow measurement was realized by the Vau method (CARVALHO, 2008). Such methodology can be used whenever the water level is below 1.2 meters.

Tributary T2 is the Leme Stream which is channeled along its extension and its source is located in the city center. On its way to Igapó I Lake, it receives domestic sewage discharge from the whole area. T2 flow measurement was carried out by the salt dilution method (COMINA et al., 2013).

Tributary T3 includes a source and a drain. The source is a stream which has been channeled through circular pipes of 1 meter of diameter. The open-ended drain is a rectangular spillway; therefore, the measurements of T3 were realized by two types of calculations. The Mannig-Stricker equation (RAJU, 1981) was used for the stream that runs into Igapó I Lake while in the drain, flow was calculated through a spillway equation (SIMÕES, 2013).

In addition to these distinct methodologies, three campaigns for the tributaries flow measurement were developed in days under regular and normal conditions. Campaign results are presented in Table 1 and they show that the largest flow was obtained during the first campaign, with a contribution of 1095 L/s. The third campaign was measured after a rainy period; however, only tributary T2 showed the largest flow. The tributaries contributions mean was of 946 L/s.

**Table 1:** Tributaries flow in L/s

| Tributary | 1$^{st}$ Campaign | 2$^{nd}$ Campaign | 3$^{rd}$ Campaign | Flow Means |
|---|---|---|---|---|
| T1 | 1043.0 | 838.9 | 690.5 | 857.5 |
| T2 | 38.8 | 57.9 | 132,6 | 76.5 |
| T3 | 13.2 | 10.0 | 13.0 | 12.0 |
| Total | 1095 | 906.8 | 836.1 | 946.0 |

**Source:** the authors themselves.

According to results presented in Table 1, the flow into Igapó I Lake is strongly affected by tributary T1, since it contributes with approximately 91% of the effluents discharged into Igapó I Lake. Tributaries T2 and T3 contribute with 8% and 1% of the effluents discharged into Igapó I Lake, respectively.

**Computational model**

In this chapter, we first describe how the computational mesh was generated, and then we describe the hydrodynamic and BOD$_5$ transport models. The numerical procedures used are detailed as well. In the last section, the methodology used for the calibration of the deoxygenation coefficient is presented.

Throughout the chapter the computational model used the same physical domain scale of Igapó I Lake. Therefore, the following was taken into consideration: the real geometry of the lake, the localization of water sample collection points, the respective BOD$_5$ concentration values analyzed in the laboratory and the dynamics of the water circulation flow (ROMEIRO et al., 2011; PARDO et al., 2012).

**Computational mesh generation**

Domain discretization was obtained by using structured mesh due to their facility in ordering elementary elements. In addition, curvilinear coordinates were also adopted since this system allows the meshes to coincide with the geometry of the problem, making computational treatment more adequate. Other reasons that led to the use of the curvilinear coordinates system in mesh discretization were its easy computational code programming process and facility to develop generic methodologies (MALISKA; RAITHBY, 1984;

CIRILO; BORTOLI, 2006). Moreover, the discretized mesh in curvilinear coordinates, whose elements are quadrilateral, was easily adapted to be used in the finite elements code.

Thus, using satellite images from the app Google Earth (GOOGLE EARTH, 2016), it was possible to collect data from the lake's contours and tributaries T1, T2 and T3 which run into Igapó I Lake. Data presented in Table 1 show that the largest water flow into the Igapó I Lake comes from tributary T1 and; therefore, in the numerical simulations the effects caused by tributaries T2 and T3 will not be considered.

About the computational mesh generation, the right and left margins of the lake were obtained by a parameterized polynomial cubic spline interpolation (THOMPSON; WARSI; MASTIN, 1985; CIRILO; BORTOLI, 2006; FANG; HUNG, 2013). In regards to the coordinated lines in the interior of the computational mesh, they are generated by the standard elliptic EDPs system (MALISKA; RAITHBY, 1984; ROMEIRO et al., 2011).

**Hydrodynamic model**

The mathematical modelling of the problem was established for a Newtonian, incompressible, laminar and isothermal flow so that density and viscosity are held constant. It was also understood that the lake contours do not change and that the action of the wind on the water mirror is not significant. These hypotheses are consistent with the Igapó Lake flow that presents low speed waters, due to a small entry volume, when compared to the volume of the reservoir and to the lake's low declivity. In addition, the Londrina region is characterized by calm winds, including the region of the water mirror of the Igapó I Lake, except under atypical meteorological conditions, which are not part of this work's scope.

Taking these hypotheses, the two-dimensional hydrodynamic model, in the dimensionless form, is given by

$$\frac{\partial u}{\partial t} + u\frac{\partial u}{\partial x} + v\frac{\partial u}{\partial y} = -\frac{\partial p}{\partial x} + \frac{1}{Re}\left(\frac{\partial^2 u}{\partial x^2} + \frac{\partial^2 u}{\partial y^2}\right) \qquad (1)$$

$$\frac{\partial v}{\partial t} + u\frac{\partial v}{\partial x} + v\frac{\partial v}{\partial y} = -\frac{\partial p}{\partial y} + \frac{1}{Re}\left(\frac{\partial^2 v}{\partial x^2} + \frac{\partial^2 v}{\partial y^2}\right) \qquad (2)$$

$$\nabla^2 p = -\frac{\partial^2 uu}{\partial x^2} - 2\frac{\partial^2 uv}{\partial xy} - \frac{\partial^2 vv}{\partial y^2} - \frac{\partial d}{\partial t} + \frac{1}{Re}\left(\frac{\partial^2 d}{\partial x^2} + \frac{\partial^2 d}{\partial y^2}\right) \qquad (3)$$

where $t$ is the time, $u$ and $v$ are the components of the speed vector in the $x$ and $y$ directions, respectively, $Re$ is the Reynolds number, $d = \frac{\partial u}{\partial x} + \frac{\partial v}{\partial y}$ is the speed divergent and $p$ is the pressure.

An algebraic system of equations is obtained by rewriting equations (1) − (3) in the system of curvilinear coordinates and by approximating the partial derivatives by finite differences (QUEIROZ et al., 2006; CIRILO et al., 2010). By the Succesive Over Relaxation Method (CIRILO et al., 2008), the algebraic Poisson's equation originating from (3) was resolved iteratively, generating the pressure field for the whole computational domain. The velocity components ($u$ and $v$) were obtained for the Igapó I Lake geometry by inserting the pressure field in the algebraic equations (1) and (2), and these equations were solved via Runge-Kutta Method of third order (ROMEIRO et al., 2011; PARDO et al., 2012). The MAC (Marker-and-Cell) Method was the technique used to solve the system of equations (AMSDEN; HARLOW, 1970). The velocity field obtained by the procedure will be considered as entry data in the transport model.

**BOD$_5$ transport model**

In lakes without a high concentration of suspended sediments, the reactive species, dissolved in the water body, flow with the velocity field of the lake. In these situations, the reactive species are in passive regime, and the study of their transport can be carried out independently of the hydrodynamic modeling. In the context of Lake Igapo, we consider a mathematical modeling in which the hydrodynamic model is decoupled from the transport model. Thus, the BOD$_5$ transport simulation was carried out by an advection-dispersion-reaction two-dimensional model given by

$$\frac{\partial C}{\partial t} + u\frac{\partial C}{\partial x} + v\frac{\partial C}{\partial y} - D_x \frac{\partial^2 C}{\partial x^2} - D_y \frac{\partial^2 C}{\partial y^2} = -K_1 C \qquad (4)$$

where $C = C(x, y, t)$ represents the BOD$_5$ concentration, $u$ and $v$ are the speed obtained by the hydrodynamic model, $D_x$ and $D_y$ are the longitudinal and transversal diffusion coefficients and $K_1$ is the BOD$_5$ deoxygenation coefficient.

The Semi-Discrete Stabilized Finite Element, a method that combines finite differences in time and finite elements in space (ROMEIRO et al., 2007; MALTA; CASTRO, 2010, LADEIA, C. A. et al, 2013), was used to resolve the model presented in eq. (4) under a null initial condition and a frontier condition given by $C = C_o(t)$.

The computational mesh adopted by the transport model was the same that generated the speed fields, considering that it was adapted to be used with the finite elements code in which the elements are triangular.

For the $K_1$ deoxygenation coefficient, the numerical model took into account the values measured at four points in the Igapó I Lake, during the months of February and March, 2013.

**$K_1$ deoxygenation coefficient calibration**

For the transport model simulation, eq. (4), it is necessary to obtain the $K_1$ deoxygenation coefficient, considered constant in the model. The $K_1$ value can be calculated from the BOD$_5$ value by the Thomas method (THOMAS, 1950). In this context, water samples were collected at four points along the lake, represented by P1, P2, P3 and P4, shown in Figure 2.

Table 2 shows deoxygenation coefficient values for each P point, using the Thomas method (THOMAS, 1950). The $K_1$ values range presented in Table 2 indicates the existence of sewage discharge points in the Igapó I Lake tributaries (VON SPERLING, 1996)

**Table 2:** $K_1$ values for the four analyzed samples

|  | P1 | P2 | P3 | P4 |
|---|---|---|---|---|
| $K_1 (d^{-1})$ | 0.445 | 0.424 | 0.478 | 0.447 |

**Source:** the authors themselves.

**Figure 2:** Satellite image of Igapó I Lake with the locations of the four P points used for $K_1$ deoxygenation coefficient calibration.

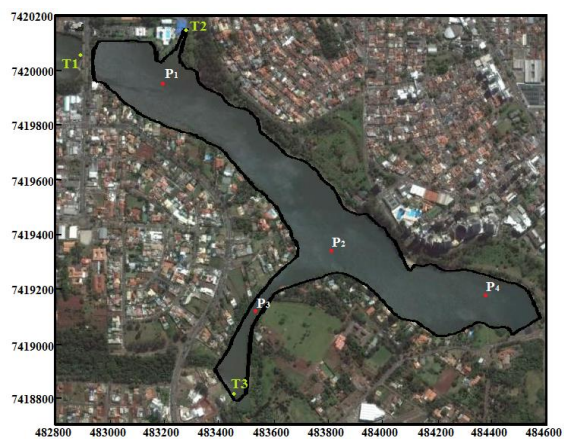

**Source:** Google Earth, 2016 (adapted).

**$BOD_5$ concentration verification points**

Considering the $K_1$ values obtained in Table 2, it becomes possible to evaluate the quality of the water from the numerical model, having as parameter the BOD5 analysis values. The Table 3 shows $BOD_5$ concentration values measured in 5 points represented by Li, $i = 1, 2, \ldots, 5$, where the points $x$ and $y$ meet in UTM (Universal Transverse Mercator) coordinates, obtained in February 2013.

In addition to the L points described in Table 3, Figure 3 shows $r = 25\,m$ radius circles positioned around each L point. These circles were included as an alternative to evaluate $BOD_5$ simulated concentrations values next to verification points.

**Table 3:** $BOD_5$ analyses values, according to locations.

| $x$ | $y$ | Locations | $BOD_5$ (mg/l) |
|---|---|---|---|
| 482935 | 7420043 | $L_1$ | 12.6 |
| 483239 | 7419906 | $L_2$ | 10.8 |
| 484077 | 7419213 | $L_3$ | 7.6 |
| 484406 | 7419146 | $L_4$ | 7.0 |
| 484528 | 7419065 | $L_5$ | 6.5 |

**Source:** the authors themselves.

**Figure 3:** Satellite image of Igapó I Lake showing the five L points used to verify $BOD_5$ concentrations coefficients.

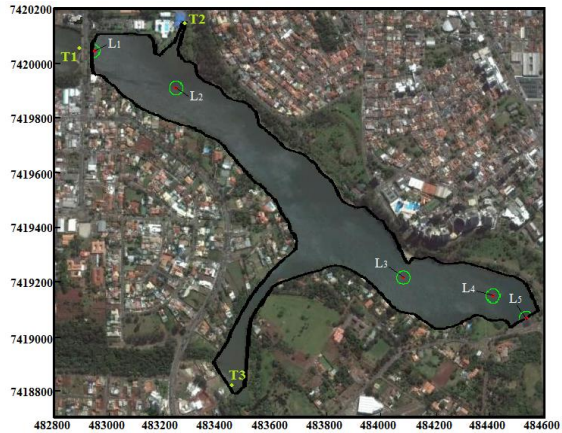

**Source:** Google Earth, 2016 (adapted).

**Results and discussion**

By taking as entry data for the transport model, eq. (4), the mesh generated by the curvilinear coordinates and the speed fields generated by the hydrodynamic model, it was possible to obtain, from the numerical simulations, the field of $BOD_5$ concentrations for the whole lake.

To generate the interior of Igapó I Lake geometry, a structured discretization was used, in generalized coordinates, whereas the margins were obtained by a parametric cubic spline polynomial interpolation. Such procedures were adopted due to their computer performance as well as by the fast similarity obtained with the physical geometry, from little known points of the domain (MALISKA, 1995). The study considered 839 points located along the left and right margins and 35 points located in the entrance and exit contours. To solve the resulting tridiagonal linear systems, the TDMA (TriDiagonal Matrix Algorithm) procedure was utilized, which reduces the memory time.

To generate the velocity field, $Re = 10,0$ (Reynolds number) was used, which is consistent with the T1 flow, given in Table 1. Thus, the following boundary conditions are considered: condition prescribed at the input given by $Re = 10.0$, free output condition and non-slip condition at the border (fluid does not flow at the border). To optimize the convergence, we used $w = 0,7$ in the SOR method (CIRILO et al., 2008) and tolerance of $10^{-4}$. The PDEs systems (1-3) were resolved in generalized coordinates. For the discretized velocity field was used the explicit third order Runge-Kutta method and for the discretized pressure field was used the Gauss-Seidel method with successive relaxations. The MAC (Marker-and-Cell) Method was the technique used to solve the system of equations.

Regarding to the transport model parameters the following were taken: for the longitudinal and transversal diffusion coefficients $D_x = D_y = 0,001$ $m^2 d^{-1}$ (ROMEIRO et al., 2011), and for deoxygenation coefficients values, $K_1$ in $d^{-1}$, shown in Table 2. Regarding the boundary conditions, the BOD5 concentration at the T1 entry was taken as 12.6 mg / l, this value consistent with the experimental measurements. Finally, the simulations were carried out for the total time of 10 days with $\Delta t = 0,5\,d$. For the numerical resolution of the transport model (4), the stabilized finite elements method was employed, in its Galerkin's semi-discrete formulation, when the spatial derivate is approached by finite elements and temporal derivate is approached by finite differences. A stabilization procedure of the Streamline Upwind PetrovGalerkin (SUPG) type, proposed by Brooks and Hughes (1982), was also employed.

### $K_1$ Simulations

The purpose of this subsection is to simulate the best $K_1$ value for Lake Igapo I. Simulations of the numerical model were realized considering the four $K_1$ deoxygenation coefficients, calculated at the P points, Table 2, obtaining $BOD_5$ concentrations for the whole lake. Next, the numerical results were compared to 5 $BOD_5$ values measured at the L points, shown in Figure 3. Next, only 10 points were selected inside the circumference, $r \leq 25\,\text{m}$, whose $BOD_5$ concentration values were closer to the values measured experimentally – see Table 3. With this comparison, it was possible to assess which $K_1$ value reproduces numerical solutions closer to reality.

During the analysis to determine $K_1$, $BOD_5$ simulated values for all points in the $r \leq 25\,\text{m}$ radius circumference, positioned around each L point were verified. These results are presented in Figure 4, which shows, in the first column, how far the best $BOD_5$ simulated values are from their respective $BOD_5$ values measured experimentally. The second column, in Figure 4, shows how far the best $BOD_5$ simulated values are from their respective L points.

Still, first column in Figure 4, the L2 point shows great $BOD_5$ variation around the experimental value, for all tests. This variation is justified by the fact that L2 is found between two great vortices right after the lake entrance (PARDO et al., 2012). For the evaluations of the other locations, the model showed results quantitatively satisfactory with the real behavior of the Igapó I Lake in relation to the simulated parameter.

In regards to the numerical results presented in the second column of Figure 4, it was possible to understand how the simulated results for $K_1$ affected the $BOD_5$ concentrations analyzed in relation to radiuses of 5, 10, 15, 20 and 25 meters of distance from the measured points. Figure 4 also shows that $K_1 = 0{,}478\,\text{d}^{-1}$ generated numerical solutions which were closer to the lake reality.

### $BOD_5$ Simulations

Taking $K_1 = 0{,}478\,\text{d}^{-1}$, the $BOD_5$ concentrations map for the whole Igapó I Lake is presented in Figure 5. Simulation time was of 10 days, when the permanent regime for the $BOD_5$ concentrations has already been achieved.

Figure 5 presents the five L points and their respective $BOD_5$ values. It also shows the vortices next to the $L_2$ location, making the simulations variations in this location more comprehensible. Moreover, based on the simulated results, the $BOD_5$ concentration was reduced from 12.6 mg/L to 6.5 mg/L, showing satisfactory quantitative results with the real behavior of Igapó I Lake. Finally, it also showed that the numerical model was not sensitive to cuts in the geometry by excluding tributaries T2 and T3.

**Figure 4:** Simulated results for the 10 points whose BOD$_5$ concentration values are closer to the measured values. a) P1 point; b) P2 point; c) P3 point; and d) P4 point.

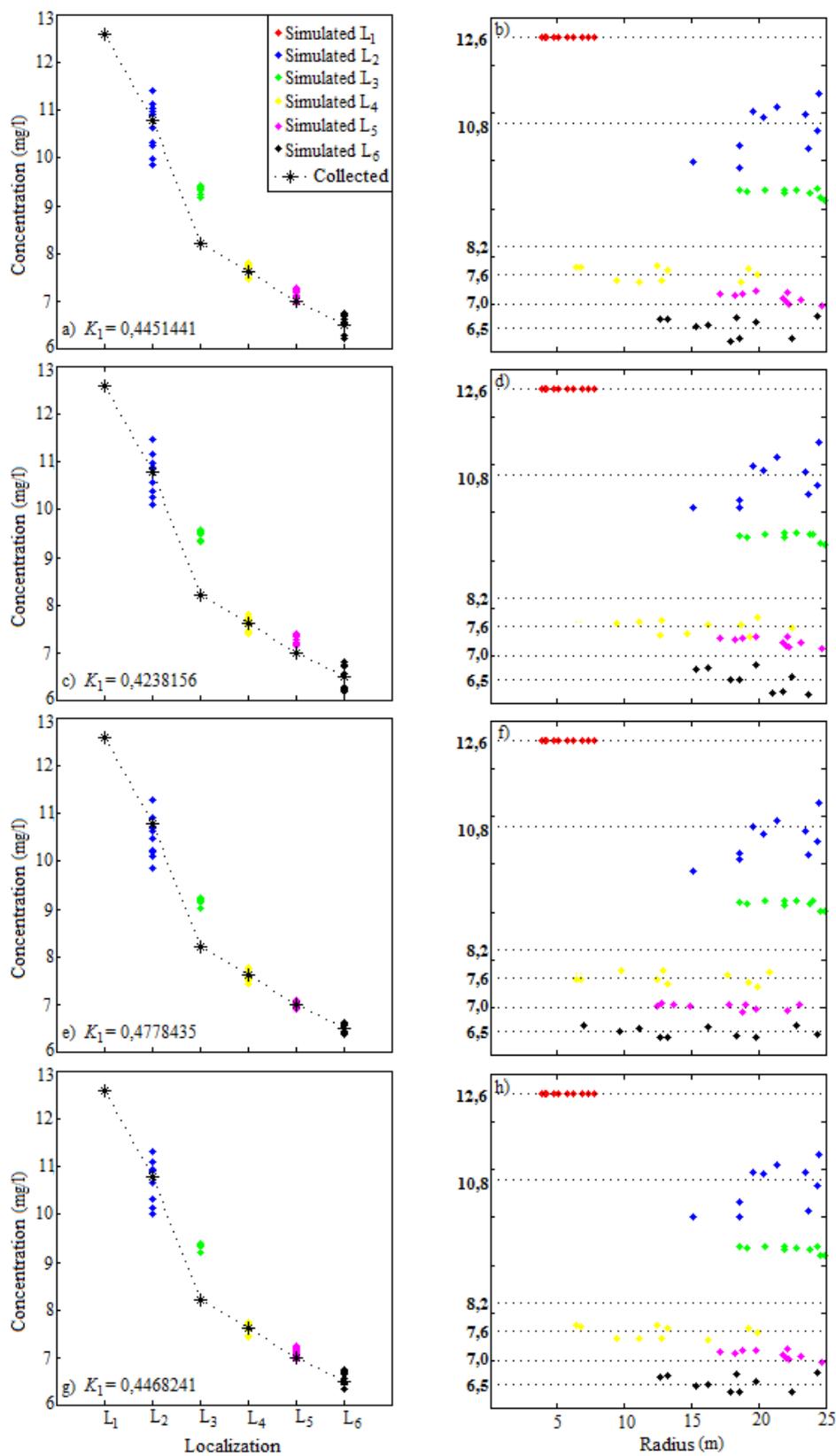

**Source:** the authors themselves.

**Figure 5:** Simulated values of BOD$_5$ concentration, in the permanent regime, for Igapó I Lake.

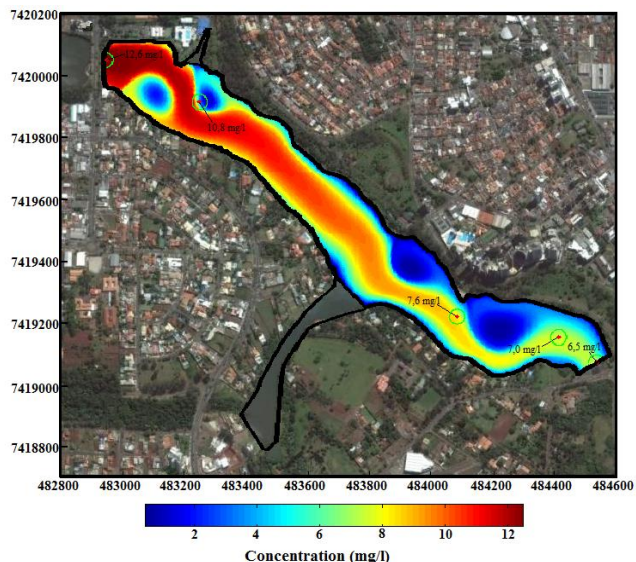

**Source:** the authors themselves.

**Final Considerations**

CONAMA (Brazilian National Environment Council) recommend that the primary contact recreation and sports practices occur when the BOD$_5$ is less than 5 mg/L. The results observed and simulated to BOD$_5$ concentrations in Igapó I Lake presented values above those allowed by Brazilian Legislation.

By observing the BOD$_5$ concentration in the left entrance of the lake, where the Iate Club promotes sports practices, and the concentrations on the right side, where there are residences whose owners use the lake for recreation, one sees that the BOD$_5$ concentrations are higher than 10 mg/L. The exit of the lake has a physical dam and the water flow is controlled by water pipes and ramps, which in the summer period are used by the community for recreation. Analogously, the simulations also show that in this region the BOD$_5$ concentration is higher than the CONAMA recommendation. Moreover, the high BOD$_5$ concentration causes a drop in dissolved oxygen (OD) concentration, and consequently of other parameters, damaging the aquatic biodiversity of Igapó I lake. Therefore, based on our simulations, we do not recommend primary contact with the waters of the lake, and that does not allow fishing by the population.

In this context, the use of mathematical modeling, calibrated and validated, is a useful and important tool for assessing water quality in water bodies, presented results that can be adopted by public officials for the recovery of the water quality in Igapó I Lake, Londrina, Paraná, Brazil.

**Acknowledgements**

The authors are grateful for the financial support of Coordenação de Aperfeiçoamento de Pessoal de Nível Superior (CAPES), publication 24/2012 - Pró-Equipamentos Institucional, agreement 774568/2012. The author N.M.L. Romeiro acknowledges Fundação

Araucária (FA) for the partial financial support to this research (FA 39225/24-2012), grants 725/2013.